\documentclass[a4paper]{article}

\usepackage[margin=1in]{geometry} 

\usepackage{amsmath}
\usepackage{amsthm}
\usepackage{amssymb}

\usepackage[utf8]{inputenc}

\usepackage{graphicx, color}
\graphicspath{{fig/}}

\title{Shearless curve breakup in the biquadratic nontwist map}
\author{G. C. Grime$^1$\thanks{E-mail: gabrielgrime@gmail.com} \and M. Roberto$^2$ \and R. L. Viana$^3$ \and Y. Elskens$^4$ \and I. L. Caldas$^1$}

\date{
	$^1$Institute of Physics, University of São Paulo, São Paulo, 05508-090, Brazil\\
	$^2$Physics Department, Aeronautics Institute of Technology, São José dos Campos, 1228-900, Brazil \\
	$^3$Physics Department, Federal University of Paraná, Curitiba, 81531-990, Brazil\\
	$^4$Aix-Marseille Université, UMR 7345 CNRS, PIIM, Marseille, cx 13, 13397, France\\
	\today
}

\begin{document}
	\maketitle
	
	\begin{abstract}
        \noindent Nontwist area-preserving maps violate the twist condition along shearless invariant curves, which act as transport barriers in phase space. Recently, some plasma models have presented multiple shearless curves in phase space and these curves can break up independently. In this paper, we describe the different shearless curve breakup scenarios of the so-called biquadratic nontwist map, a recently proposed area-preserving map derived from a plasma model, that captures the essential behavior of systems with multiple shearless curves. Three different scenarios are found and their dependence on the system parameters is analyzed. The results indicate a relation between shearless curve breakup and periodic orbit reconnection-collision sequences. In addition, even after a shearless curve breakup, the remaining curves inhibit global transport.	\par
	\end{abstract}

\section{Introduction}
\label{sec:introduction}

The twist condition plays an important role in Hamiltonian systems due to its connection with important results, e.g., the Kolmogorov-Arnold-Moser theorem, Poincaré-Birkhoff theorem, Aubry-Mather theory, and Nekhoroshev theorem \cite{meiss1992,lochak1992}. The implications of this condition to the system dynamics range from the number of periodic orbits to its transport behavior \cite{lichtenberg}. The violation of the twist condition generates new phenomena that significantly change the system behavior \cite{morrison2000}. The so-called nontwist systems present shearless invariant curves that violate the twist condition. They are invariant curves resistant under periodic perturbations, and they are also called shearless transport barriers \cite{caldas2012}.

The Standard Nontwist Map (SNM) is a paradigmatic area-preserving map in the study of nontwist phenomena close to the shearless curve \cite{diego1996}. It is widely studied in the context of reconnection-collision sequences \cite{petrisor2001,wurm2004,wurm2005,shinohara1998}, shearless curve breakup \cite{diego1997,apte2003,apte2005,mathias2019}, and transport properties \cite{szezech2009,mugnaine2018,viana2021}.

Physical systems with a nonmonotonic profile can have nontwist dynamics. Examples of such systems include: super-conducting quantum interference devices \cite{soskin1994}, traveling waves \cite{weiss1991}, sheared zonal flows \cite{diego1993,diego2000}, and magnetically-confined plasmas \cite{caldas2012,horton1998}. In these systems, especially in fluids and plasmas, the shearless transport barrier resembles a transport barrier in the real physical system \cite{caldas2012,diego2000}. For example, experimental observations show a correlation between nonmonotonic electric or magnetic field profiles in tokamak plasma and the reduction of transport \cite{connor2004}.

Many methods are used to evaluate the critical parameters associated with the shearless curve breakup. One of them is based on Greene's criterion \cite{greene1979}, modified to nontwist systems \cite{diego1996}. Although very precise, its computation cost may restrain its use to compute parameter spaces. Meanwhile, other methods produce reliable results in this task, e.g., methods based on the escape of trajectories \cite{shinohara1997}, the convergence of rotation number \cite{wurm2005}, Slater's criterion for the existence of quasiperiodic orbits \cite{abud2015}, and recurrence-based analysis \cite{santos2018}.

Some investigations concerning plasma-transport models have nontwist systems that admit multiple shearless tori in phase space \cite{martinell2013,grime2023JPP,osorio2021}. Also, experimental evidence indicates that physical systems can present multiple shearless transport barriers \cite{joffrin2003}. Despite some works in the understanding of these systems \cite{howard1995,dullin2000,grimeBNM}, their shearless curve breakup scenarios and the corresponding transport behavior are still an open question. In addition, a recent paper proposed a new area-preserving map, derived from a plasma-transport model, that has three shearless curves \cite{grimeBNM}. This system, called Biquadratic Nontwist Map (BNM), presents shearless bifurcation scenarios with the same characteristics encountered in the original Hamiltonian flow from which it was derived \cite{grime2023JPP}. This area-preserving map has connections with the SNM and shares the same symmetry and involution properties which simplifies analytical and numerical calculations \cite{petrisor2001,grimeBNM}.

In this work, we progress in the study of the BNM, characterizing its shearless curve breakup scenarios and relation with the system parameters. We determine the shearless curve breakup parameter spaces concerning the breakup of each shearless curve. The structures of the parameter space are described along with their relation with periodic orbits reconnection collision sequences. We also investigate the effects of shearless breakup in the phase space transport, through the computation of the transport barrier transmissivity.

This paper is organized as follows. The Biquadratic Nontwist Map is introduced in Sec. 2 together with its shearless curve breakup scenarios. In Sec. 3 we discuss the dependence of shearless curve breakup configuration on the system parameters. Sec. 4 concentrates on discussing the smooth boundaries in the parameter space and its relation with the reconnection of separatrices. The shearless partial transport barrier transmissivity is presented in Sec. 5. Conclusions are presented in the last section.

\section{The Biquadratic Nontwist Map}

The Biquadratic Nontwist Map (BNM) was proposed in Ref. \cite{grimeBNM}, and is given by
\begin{subequations}
\label{eq:bnm}
\begin{align}
    x_{n+1} &= x_n + a \left( 1-y_{n+1}^2 \right)\left( 1-\epsilon y_{n+1}^2 \right)\  (\mathrm{mod}\ 1)\\[0.2cm]
    y_{n+1} &= y_n - b \sin{(2\pi x_n)},
\end{align}
\end{subequations}
where $x \in [0,1)$ and $y\in \mathbb{R}$ are the angle-action like coordinates. The parameters of the map and their ranges are $a \in [0,1)$, $b \in [0,\infty)$, $\epsilon \in [0,\infty)$. Parameter $b$ is related to the perturbation of the map, and $a$ and $\epsilon$ control its twist function. In real physical systems, as in a tokamak plasma, the perturbation parameter can be related to the amplitude of perturbing waves, and the twist function, to spatial profiles, such as the velocity of a fluid or electromagnetic fields in a plasma. A more specific relation of the parameters of the map to a physical model is available in Ref. \cite{grimeBNM}.

This map violates the twist condition, $|\partial x_{n+1}/\partial y_n|\neq 0$, for three sets of points $\mathbf{z}=(x,y)$ in phase space, called shearless curves. These curves, as any invariant torus, are total barriers for transport in phase space \cite{mackay1984}. Nonetheless, they are particularly robust under periodic perturbations, and, even after their breakup, a partial transport barrier is still present \cite{szezech2009}.

\begin{figure}[htb]
    \centering
    \includegraphics[width=0.45\textwidth]{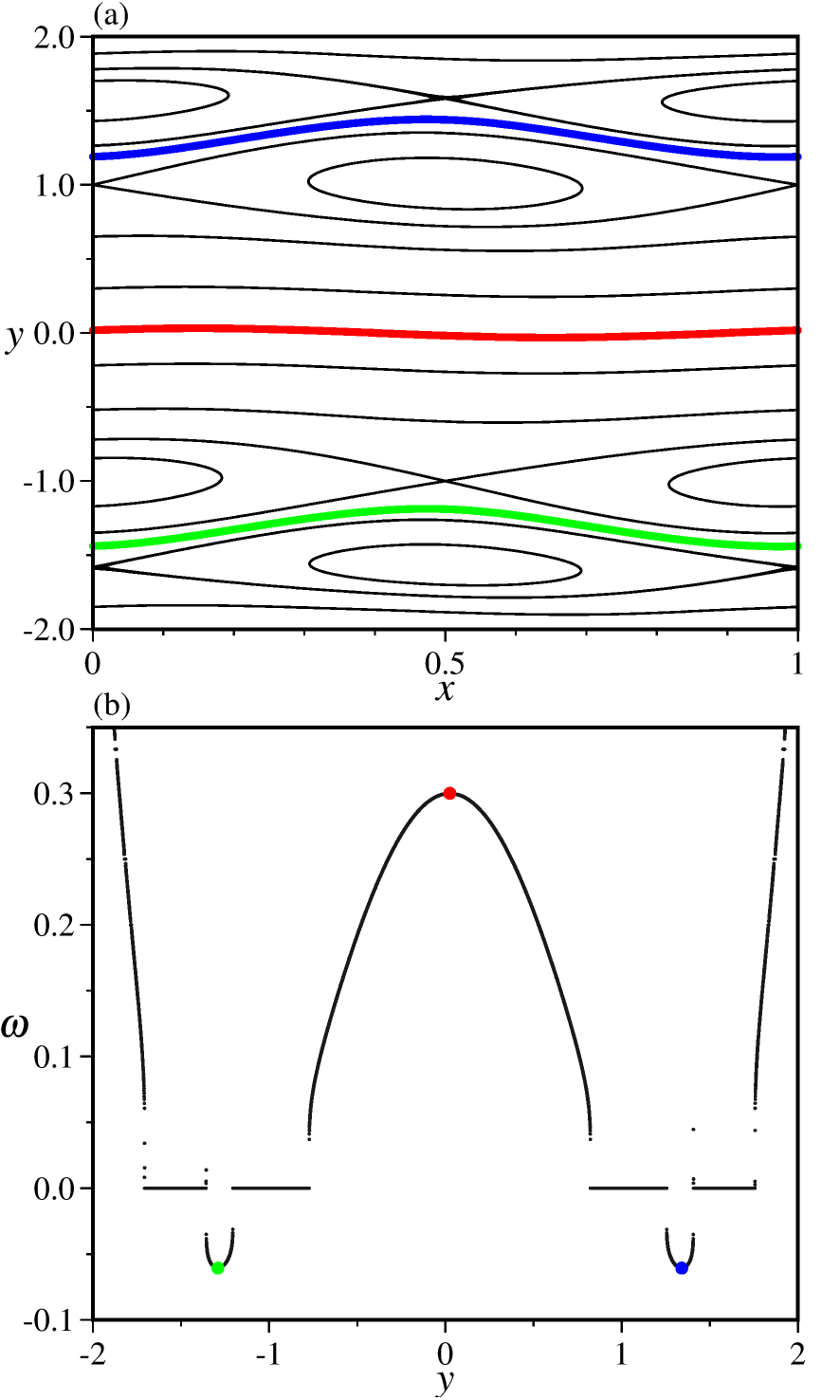}
    \caption{(a) Phase space of the Biquadratic Nontwist Map for parameters $a=0.3$, $b=0.05$, and $\epsilon=0.4$. The associated rotation number profile is shown in (b) using the initial angle $x_0=0.25$. The shearless curves are marked in red, blue and green color.}
    \label{fig:1}
\end{figure}

The map has three shearless curves, $C_1, C_2 \ \text{and} \ C_3$, which for the condition $b \ll 1$ are approximately given by the expressions
\begin{subequations}
\begin{align}
&C_1: \ y=b \sin{(2\pi x)}, \\[0.2cm]
&C_{2,3}: \ y= \pm \sqrt{\dfrac{1+\epsilon}{2\epsilon}} + b \sin{(2\pi x)},
\end{align}    
\end{subequations}
shown in Figure \ref{fig:1}a, colored in red, blue, and green, respectively. To obtain the shearless curves orbits, we numerically determine the rotation number profile, that associates with each nonchaotic orbit a rotation number
\begin{equation}
    \omega(x_0,y_0) = \lim_{m\to \infty}  \dfrac{x_{m+1} - x_0}{m},
\end{equation}
which measures the average angle variation per time step. Then, the shearless orbit is characterized by an extremum in the rotation number profile. The corresponding rotation number profile of this map is plotted in Figure \ref{fig:1}b, showing the three extreme points associated with the shearless tori.

The fixed points of the BNM are the solutions of the equation $M(x,y) = (x,y)$, where $M$ is the mapping defined by Eq. \eqref{eq:bnm}. They are given by
\begin{subequations}
\label{eq:fixed.points}
\begin{align}
\mathbf{z}_1^{\pm}&=(0,\pm 1), \hspace{0.75cm} \mathbf{z}_2^{\pm}=\left( 0,\pm \dfrac{1}{\sqrt{\epsilon}} \right), \\[0.1cm] 
\mathbf{z}_3^{\pm}&=\left(\dfrac{1}{2}, \ \pm 1\right), \hspace{0.5cm} \mathbf{z}_4^{\pm}=\left(\dfrac{1}{2}, \ \pm \dfrac{1}{\sqrt{\epsilon}}\right),
\end{align}
\end{subequations}
and are related to the four main islands and four hyperbolic points of period one (Figure \ref{fig:1}a).

In this paper, we call \textit{central main islands} those associated with the fixed points $\mathbf{z}_1^-$ and $\mathbf{z}_3^+$, also present in the SNM. The islands of the points $\mathbf{z}_2^+$ and $\mathbf{z}_4^-$ are named \textit{external main islands} and are typical of the BNM. In addition, there are two possible reconnection processes involving those four main islands. The first one occurs involving the central main islands and has an equivalent in the SMN \cite{diego1996}. Furthermore, the BNM has another reconnection that involves the central and the external main islands \cite{grimeBNM}.

Another important quantity associated with area-preserving maps is the primitive function, a generalization of the generating function of canonical transformations \cite{haro2000}. The primitive function $W(x,y)$ of the BNM is given by
\begin{equation}\label{eq:primitive}
\begin{split}
    W(x,y) &= -\dfrac{2a(1+\epsilon)}{3}\left[ y-b\sin{(2\pi x)} \right]^3 +\\
    &+\dfrac{4a\epsilon}{5}\left[ y-b\sin{(2\pi x)} \right]^5  + \dfrac{b}{2\pi}\cos{(2\pi x)},
\end{split}
\end{equation}
which, for $\epsilon=0$, reduces to the primitive function of the Standard Nontwist Map \cite{wurm2004}. Furthermore, this function can be used to find the reconnection threshold of hyperbolic orbits \cite{petrisor2002,apte2006}.

The BNM has special properties like spatial symmetry and can be decomposed in involutions, discussed in \ref{sec:appendix1}. These properties simplify some numerical procedures such as finding periodic orbits \cite{diego1996}. One of its consequences is the presence of the indicator points, namely,
\begin{equation}\label{eq:indicator.points}
    \mathbf{P}_0^\pm =\left(\pm \dfrac{1}{4}, \pm \dfrac{b}{2} \right), \ \ \mathbf{P}_1^\pm =\left( \dfrac{a}{2} \pm \dfrac{1}{4}, 0 \right),
\end{equation}
which are points associated with the symmetry and involution transformations, and belong to the central shearless curve if it exists \cite{shinohara1997}.

\begin{figure*}[htb]
    \centering
    \includegraphics[width=0.99\textwidth]{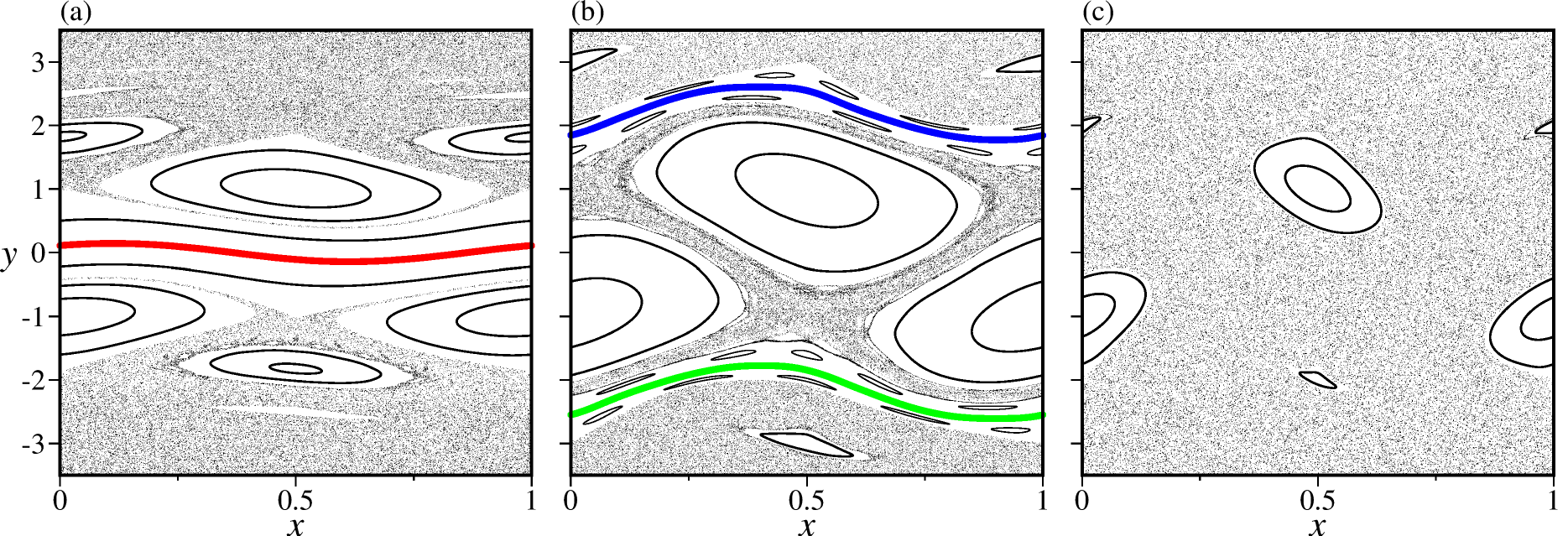}
    \caption{Shearless curve breakup scenarios of Biquadratic Nontwist Map. The phase space is shown for the parameters: (a) $a=0.23$, $b=0.19$ and $\epsilon=0.29$; (b) $a=0.135$, $b=0.57$ and $\epsilon=0.11$; and (c) $a=0.14$, $b=0.9$ and $\epsilon=0.29$.}
    \label{fig:2}
\end{figure*}

The BNM has a periodic perturbation that generates chaotic orbits starting from the hyperbolic points. The spreading of the chaotic orbits by increasing the perturbation may destroy one or more shearless curves \cite{mackay1984}. Figure \ref{fig:2} shows the three possible scenarios of shearless breakup in the BNM: (a) the central shearless curve (red) remains and the external shearless curves (blue and green) are broken; (b) the curves $C_{2,3}$ remain in phase space, but the central one is broken; and (c) all the shearless curves are broken. In the last scenario, there is no invariant curve to prevent global transport in phase space. Due to the symmetry of the map, the curves $C_2$ and $C_3$ have the same behavior, that is, if one curve is broken the other one is too. In the next section, we present a systematical analysis that finds the BNM parameter values that feature each one of the above scenarios.

\section{Shearless curve breakup in parameter space}

The breakup of the shearless curves occurs for determined critical values of the system parameters \cite{diego1996}. As mentioned in the last section, the Biquadratic Nontwist Map (BNM) can have one or more shearless curves broken. To analyze these scenarios of shearless curve breakup we obtained the shearless breakup parameter space of the Biquadratic Nontwist Map (BNM), which determines the parameters such that: (i) the central shearless curve exists, corresponding to scenarios of Figures \ref{fig:1}a or \ref{fig:2}a, or (ii) the central shearless curve is broken. In the latter case, we distinguish whether the external curves exist (Figure \ref{fig:2}b) or are broken (Figure \ref{fig:2}c).

To numerically obtain the parameters for which the central shearless curve is broken, we applied a method proposed in Ref. \cite{abud2015} based on Slater's criterion for the existence of quasiperiodic orbits \cite{slater1950,slater1967}. The method consists in counting the number of different recurrences inside a box of size $\delta$, centered at an indicator point taken as the initial condition of the orbit. If the number of different recurrences to the box is equal to three and the largest recurrence is the sum of the other two, we consider that the central shearless curve exists. Otherwise, the shearless curve is considered broken. For numerical purposes, we use the box size $\delta=0.002$ and iterate the $\mathbf{P}_0^+$ indicator point orbit $5\times 10^6$ times.

The applied method to numerically compute the parameter for which all the shearless curves are broken is the following: given a set of parameter values $(a,b,\epsilon)$ we iterate, for a long time ($10^6$), orbits with the four indicator points [Eq.\eqref{eq:indicator.points}] as the initial condition. The indicator points belong to the central shearless curve, whenever it exists. Hence, if the orbit overtakes one of the lines $y=\pm 10$, we assume that all shearless curves have broken for this set of parameters value. On the contrary, the central shearless curve exists or the external curves $C_{2,3}$ prevent the orbits from passing across the lines $y=\pm 10$. This method was extensively applied in previous works for the Standard Nontwist Map (SNM) and provides reliable results \cite{shinohara1998,mathias2019}.

\begin{figure*}
    \centering
    \includegraphics[width=0.99\textwidth]{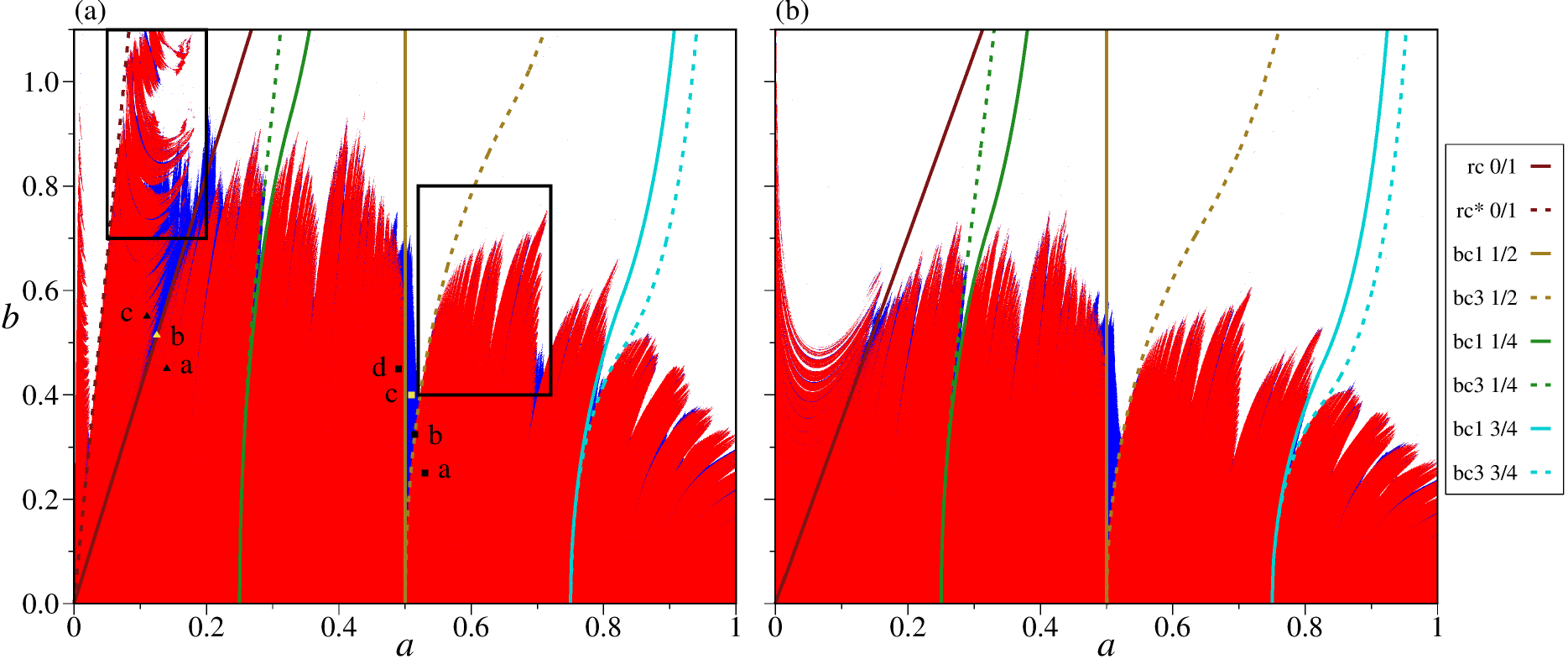}
    \caption{\label{fig:3}Parameter space of shearless curve breakup of the Biquadratic Nontwist Map, for (a) $\epsilon=0.1$ and (b) $\epsilon=0.8$. For each $(a,b)$ value, the red color represents the existence of the central shearless curve, blue the existence of the external shearless curves, while in the white region, all shearless curves are broken. The bifurcation curve of the $m/n$--periodic orbit in symmetry curve $s_j$ is denoted by bc$j$ $m/n$.}
\end{figure*}

Figure \ref{fig:3} shows the shearless curve breakup parameter space of the BNM for $\epsilon=0.1$ (Figure \ref{fig:3}a) and $\epsilon=0.8$ (Figure \ref{fig:3}b). These parameter spaces are obtained by fixing the parameter $\epsilon$ of the map \eqref{eq:bnm} and varying the parameters $a$ and $b$. Values of $(a,b)$ for which the central shearless curve exists are marked in red, in blue if the central curve is broken and the external curves exist, and in white if all shearless curves are broken.

In this paper, we will call \textit{central shearless curve parameter space} the one concerning the existence of the central shearless curve, i.e., if it exists (red region) or is broken (blue and white regions). Furthermore, the \textit{shearless curves parameter space} concerns the existence of any shearless curve (red and blue regions) or the breakup of all shearless curves (white region).

The central shearless curve parameter space present in Figure \ref{fig:3} has a mixed-type boundary combining smooth and fractal-like regions, just as in the Standard Nontwist Map \cite{wurm2005}. The fractal ones indicate that the shearless curve breakup has sensitive dependence on the system parameters. In addition, the smooth boundaries are related to the reconnection of separatrices, as we will discuss in the next section.

The red region, associated with the existence of the central shearless curve, dominates a large portion of the parameter space. However, even after the central shearless curve breakup, the external shearless curves can persist, see the blue region in the parameter space. Finally, in the white remaining portion of parameter space, all the shearless curves have broken and global transport takes place in the phase space.

Figure \ref{fig:3} shows the breakup of the central shearless curve of BNM for $\epsilon=0.1$, which is similar to the parameter space of the SNM, presented in Ref. \cite{wurm2005,mathias2019,santos2018}. In the regime of small values of $\epsilon$, the islands $\mathbf{z}_2^+$ and $\mathbf{z}_4^-$ are distant from the central region of the phase space and have a small interference in the central shearless curve breakup. This indicates that, in this regime, the SNM and the BNM have a central shearless curve with similar breakup properties. The major difference resides in the structure marked by the left box, associated with the reconnection of the external main islands of the BNM, and the SNM does not have it. More details are discussed in the next section.

For $\epsilon=0.8$ (Figure \ref{fig:3}b), the parameter space is modified. The structure associated with the external main islands does not exist and the shearless curves tend to break up for small values of the perturbation parameter $b$. In addition, the parameter space still has smooth and fractal-like boundaries.

\begin{figure}[htb]
    \centering
    \includegraphics[width=0.45\textwidth]{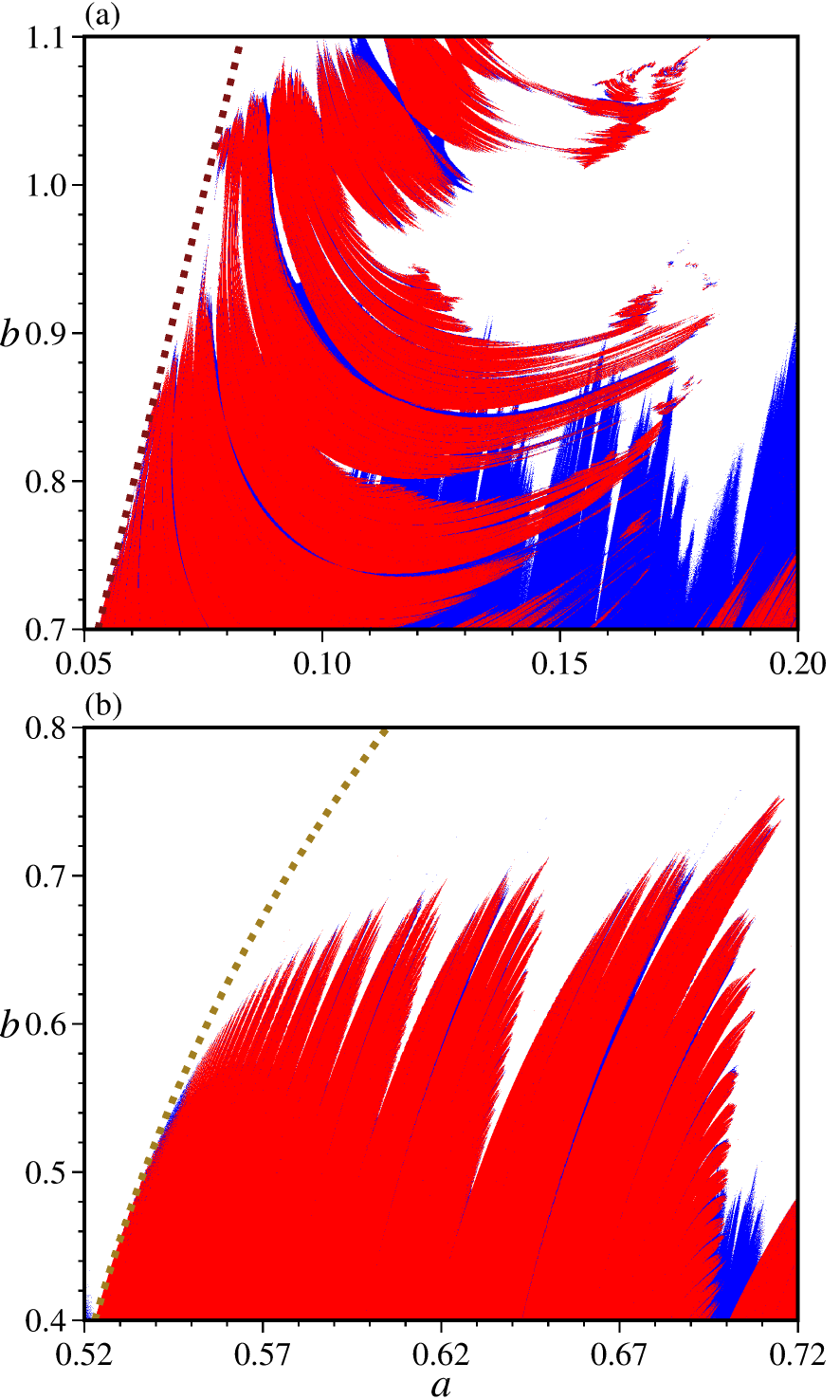}
    \caption{Magnification of the highlighted rectangles on Figure \ref{fig:3}. Red regions represent the existence of the central shearless curve, in blue the central curve is broken and the external curve persists, while in dark-green regions all shearless curves are broken.}
    \label{fig:4}
\end{figure}

Figure \ref{fig:4} shows a magnification of the rectangles highlighted in Figure \ref{fig:3}a. Especially for Figure \ref{fig:4}b, we find the same general behavior of the parameter space as in Figure \ref{fig:3}a: smooth and fractal-like boundaries. In addition, the distinction between red and blue regions is more evident.

As mentioned before, the parameter space boundary of the BNM has regions with fractal-like structures, a signature of chaotic systems. One method to quantify this fractality is based on the final state sensitivity and is known as the uncertain fraction method \cite{grebogi1985}. Small errors in the parameter value generate uncertainty in the system phase space, like the existence, or not, of the shearless curve. The numerical procedure of this method consists in calculating the fraction of uncertain points in parameter space for a given error $\delta$ and analyzing how this fraction scales with $\delta$.

More precisely, we randomly choose a large number ($10^4$) of points in the parameter space and check the existence of the central (or all) shearless curve(s). Then, we randomly choose another point $(a',b')$ in a disk of radius $\delta$ centered at $(a,b)$ and verify the shearless curve behavior. If the outcome is different from the unperturbed point, the value $(a,b)$ is considered uncertain. The fraction of uncertain points scales with $\delta$ as $f(\delta)\sim\delta^{\alpha}$, where $\alpha=2-d$ and $d$ is the dimension of the parameter space boundary \cite{grebogi1985}. More details about the application of this method to determine fractal dimensions of shearless curve breakup parameter spaces are presented in Ref. \cite{mathias2019}.

We applied the uncertain fraction method to determine the fractal dimension of the parameter space boundary in Figure \ref{fig:4}, and the results are given in Table \ref{tab:dim}. We considered two boundaries to determine the dimension: (i) the boundary of the central shearless curve breakup parameter space (red region) and the boundary of all shearless curves breakup parameter space (red and blue regions together).  The dimensions obtained confirm that the parameter space boundary has fractal dimensions close to the reported universal behavior $d=1.8$ \cite{grebogi1985}, with some deviation. The different results obtained for the fractal dimension can be related to the mixed character of the shearless curve parameter space, i.e., the boundaries have smooth and fractal portions depending on the parameter space region.

\begin{table}
    \centering
    \begin{tabular}{|c|c|c|}
        \hline
        Breakup & Fig. 4a & Fig. 4b \\
        \hline
        Central curve & $1.812 \pm 0.006$ & $1.800 \pm 0.008$ \\
        \hline
        All curves & $1.802 \pm 0.010$ & $1.772 \pm 0.009$\\
        \hline
    \end{tabular}
    \caption{Dimension of the parameter space using the uncertain fraction method. The central curve breakup refers to the boundary of the red region and the breakup of all curves denotes the red and white boundary in parameter space.}
    \label{tab:dim}
\end{table}

\

\section{Reconnection of periodic orbits}

Previous works indicate a relation between smooth boundaries in shearless curve breakup parameter space and the reconnection-collision sequences of periodic orbits \cite{wurm2004,wurm2005,shinohara1998}. In the Standard Nontwist Map (SNM), due to its symmetry, there are two standard scenarios of separatrix reconnection depending on the parity of the orbit period. In the even scenario, the separatrix reconnection is simultaneous to the collision of periodic orbits, and in the odd scenario, the collision is subsequent to reconnection \cite{diego1996}. However, in both scenarios, the separatrix collides with the shearless curve and takes its place. 

\begin{figure*}[htb]
    \centering
    \includegraphics[width=.9\textwidth]{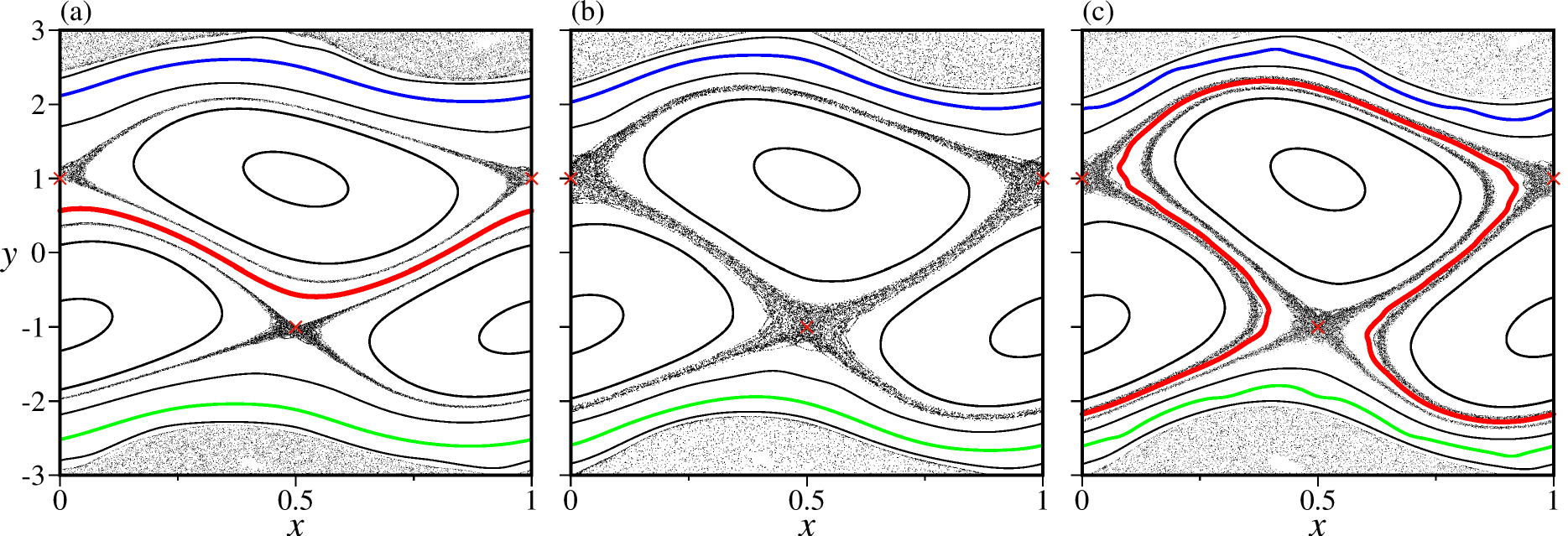}
    \caption{Central shearless curve breakup due to the reconnection of separatrices between twin islands $0/1$. Phase spaces of biquadratic nontwist map for parameters $\epsilon=0.1$, (a) $a=0.14$ and $b=0.45$, (b) $a=0.125$ and $b=0.514$, and (c) $a=0.11$ and $b=0.55$.}
    \label{fig:5}
\end{figure*}

\begin{figure*}[htb]
    \centering
    \includegraphics[width=.67\textwidth]{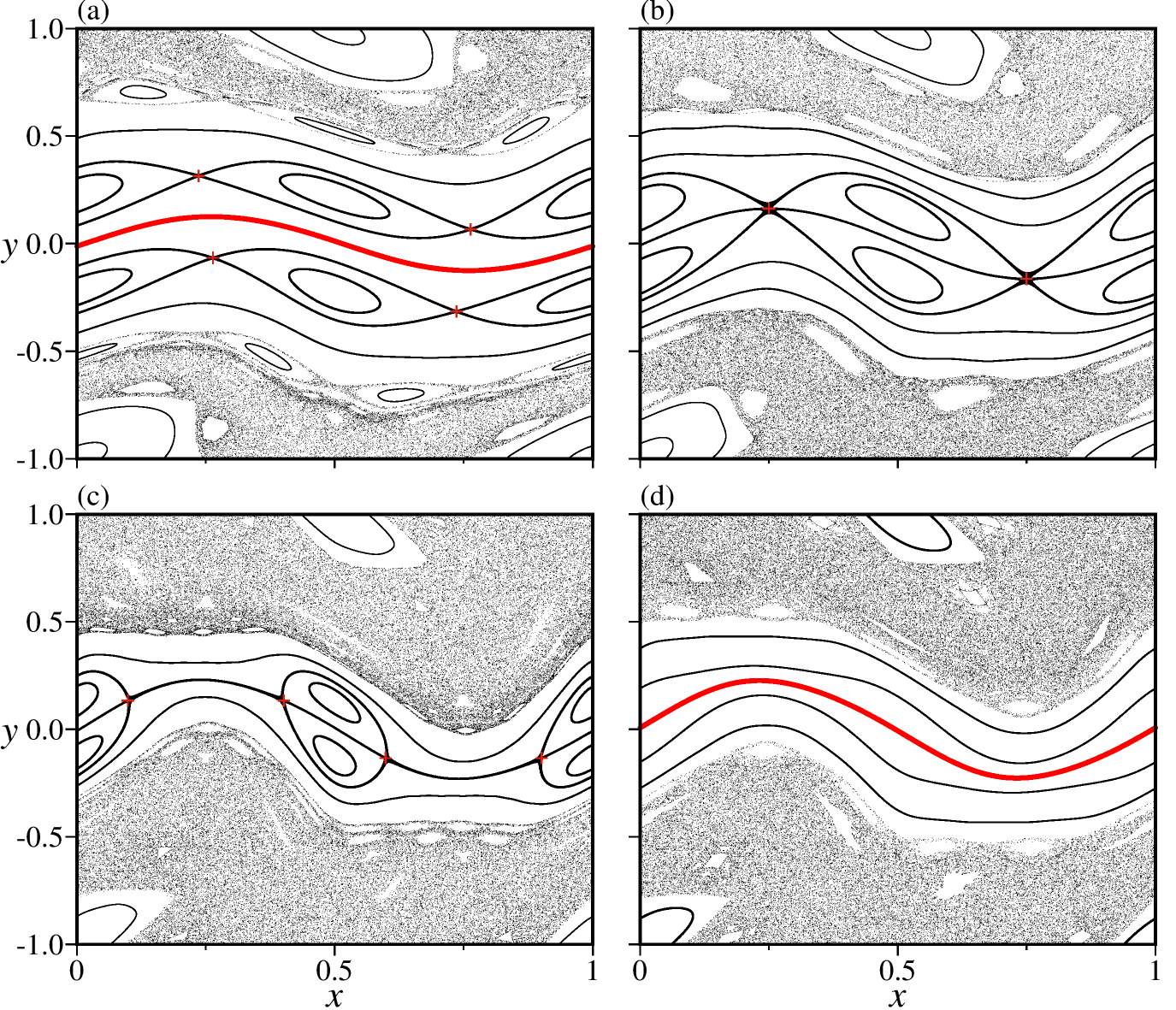}
    \caption{Central shearless curve breakup due to the $1/2$ periodic orbit collision. Phase spaces of the biquadratic nontwist map are shown for $\epsilon=0.1$, (a) $a=0.53$ and $b=0.25$, (b) $a=0.515$ and $b=0.3259$, (c) $a=0.51$ and $b=0.46$, and (d) $a=0.49$ and $b=0.45$.}
    \label{fig:6}
\end{figure*}

In the BNM, the separatrix reconnection also implies the central shearless curve breakup, as illustrated in Figures \ref{fig:5} and \ref{fig:6}. Figure \ref{fig:5} shows an odd scenario of period-one reconnection involving the central main islands, and Figure \ref{fig:6} an even scenario involving twin islands of period two. The periodic points are marked by red crosses. Both figures correspond to the parameter space points marked in Figure \ref{fig:3}a. In these two scenarios, before the reconnection, the central shearless curve exists, Figures \ref{fig:5}a and \ref{fig:6}a. At the critical parameter of separatrix reconnection, the separatrix chaotic layers collide with the central shearless curve and break it, Figures \ref{fig:5}b and \ref{fig:6}b. In the even scenario, the period two orbits (red crosses) bifurcate in two distinct periodic orbits with the same period, Figure \ref{fig:6}c. Finally, after the reconnection, the central shearless curve returns, Figures \ref{fig:5}c and \ref{fig:6}d.

To investigate the relation between separatrix reconnection and the smooth boundaries on the Biquadratic Nontwist Map (BNM) parameter space, we determine its bifurcation and reconnection curves. By definition, the bifurcation curve is the locus of points in parameter space for which two orbits of equal rotation number $\omega = q/p$, on the same symmetry line $s_j$, are at the point of collision \cite{diego1996}. Such a bifurcation curve is denoted as bc$j$ $q/p$. In addition, reconnection curves have a similar definition but are related to critical parameters for the reconnection of separatrices between twin islands \cite{wurm2004}. Finally, like in the even-period scenario, the periodic points collision occurs simultaneously with the separatrix reconnection, its bifurcation curves coincide with the reconnection curves.

To obtain the bifurcation curves for the BNM, we applied a method based on the numerical  search for the periodic orbits and the parameters for which these points collide \cite{fuchss2006phd}. The periodic orbits search used a one-dimensional root-finding thanks to the map involutions properties \cite{diego1996}, given in \ref{sec:appendix1}.

Furthermore, the reconnection threshold can be found by a criterion proposed in Ref. \cite{petrisor2002}. In this work, the author proves, for a class of standard-like maps, that hyperbolic points belonging to periodic orbits with the same period have the same action at the reconnection threshold. Hence, given the action of the BNM, Eq. \eqref{eq:primitive}, the threshold of the reconnection between the central main islands occurs when $W(\mathbf{z}_{1}^{+}) = W(\mathbf{z}_{3}^{-})$, that is, for
\begin{equation}
    b=\dfrac{4\pi a}{3}\left( 1 - \epsilon/5 \right).
\end{equation}
For a given $\epsilon$, this functional relation $b=\Psi(a)$, provides the reconnection curve rc 0/1. There is a similar reconnection curve associated with separatrix reconnection relative to fixed points $\mathbf{z}_1^+$ and $\mathbf{z}_2^+$, whose bifurcation curve is given by

\begin{equation}
    b = 2\pi a \dfrac{( -1 + 5\epsilon - 5\epsilon^{3/2} + \epsilon^{5/2})}{15\epsilon^{3/2}}.
\end{equation}
This reconnection is also related to orbits with rotation number $0/1$, so it is denoted as rc* 0/1.

In Figure \ref{fig:3}, we plot the bifurcation and reconnection curves together with the shearless curve breakup parameter space of the BNM. Indeed, the smooth boundaries of the central shearless breakup parameter space correspond, approximately, with the reconnection curves.

Typically, at the points below a specific bifurcation/reconnection curve, the periodic orbits are separated and have not collided/reconnected yet. The points marked with the letter ``a'' in parameter space represent such a situation. The points ``b'' correspond to the reconnection threshold. Finally, for the parameters above the curves, marked with the letter ``c'' in parameter space, the collision/reconnection has occurred. The phase spaces of those points are shown in Figures \ref{fig:5} (odd scenario) and \ref{fig:6} (even scenario). As we can see, the reconnection of separatrices in both scenarios leads to the breakup of the central shearless curve. In addition, in the even scenario, the reconnection is identified by the collision of the associated period orbits (marked by red points).

The reconnection of separatrices shown in Figure \ref{fig:5} involves the central main islands. This reconnection, denoted as rc 0/1, is present in both parameter spaces of Figure \ref{fig:3}. However, the BNM has a reconnection involving the external and central main islands \cite{grimeBNM}, whose reconnection curve is denoted by rc* 0/1. This reconnection curve is present only in Figure \ref{fig:3}a and is responsible for the structure marked by the left box in this Figure. In summary, the value of the $\epsilon$ parameter affects the external main island positions, which can influence the reconnection of main resonances and, consequently, modify the associated parameter space boundaries.

\

\section{Transmissivity of the transport barrier}

In the previous section, we study the shearless curve breakup parameter space of the Biquadratic Nontwist Map (BNM). That  dynamical characterization provides the parameters of the system in which one, or more, shearless curves exist or are broken. However, even after all shearless curves breakup, an effective transport barrier in phase space may persist depending on the system parameters \cite{szezech2009}. To take into account this phenomenon, dynamic quantities related to transport can be used.

The transmissivity of a transport barrier is defined as the ratio of the number of orbits that cross the barrier, over the total number of orbits. It was used before to measure the strength of the partial transport barrier of the Standard Nontwist Map (SNM) \cite{szezech2009,viana2021,mugnaine2018}. To numerically determine this quantity, we set a large number of initial conditions (we use $10^5$) on the line $\left\{(x,y): 0\le x < 1, y=-5\right\}$ (below the inferior shearless curve) and iterate each of them $5\times 10^3$ times. The transmissivity was determined as the fraction of orbits that reach the line $\left\{(x,y): 0\le x < 1, y=+5\right\}$ (above the superior shearless curve).

\begin{figure}[htb]
    \centering
    \includegraphics[width=.47\textwidth]{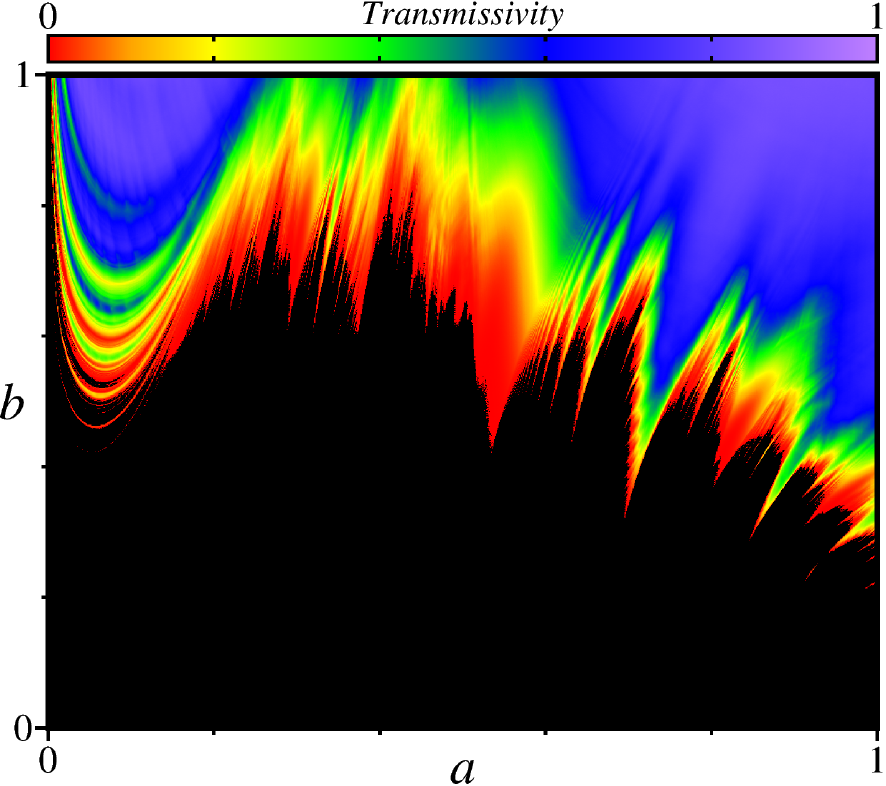}
    \caption{Transmissivity, the fraction of escaping orbits, parameter space of the Biquadratic Nontwist Map for $\epsilon=0.5$. In the black regions, there is a total barrier (shearless curves exist) and the colorbar represents the barrier transmissivity.}
    \label{fig:7}
\end{figure}

The transmissivity parameter space of the BNM, using $\epsilon=0.5$, is shown in Figure \ref{fig:7}. The colorbar represents the fraction of orbits that pass the barrier. In the black regions, this fraction of orbits is zero, that is, the system has a total barrier. The colored regions vary from red (strong partial barrier) to purple (weak partial barrier). The general behavior is similar to the result for the SNM: regions close to the shearless curve breakup have small transmissivity due to the partial transport barrier formed after the shearless curve breakup. Furthermore, in some regions, the transmissivity may increase significantly with small variations in system parameters. Other works indicate that this variation in the transmissivity is related to the resonance overlap and manifold crossing between twin islands. As mentioned in Ref. \cite{szezech2009}, small variations in the parameters of the SNM can change between intracrossing and intercrossing of the stable and unstable manifolds determining the escape channels between the islands.

\section{Conclusions}

This work presented a study of the shearless curve breakup in the Biquadratic Nontwist Map (BNM). This map, derived as an approximation of a plasma-transport model, presents three shearless curves, one central curve, and two external curves, which can break up independently in different configurations. The main contribution of this paper is to present a numerical study of the destruction of multiple shearless curves in a nontwist area-preserving map, the so-called Biquadratic Nontwist Map. The shearless curve has physical importance since it is a transport barrier, and its breakup represents a reduction in the confinement of chaotic orbits. Specifically, shearless transport barriers clarify the properties of transport and confinement in plasmas and fluid systems.

We found three different shearless breakup scenarios that influence the transport in phase space. They are: (a) only the external curves broken, (b) only the central curve broken, and (c) all the shearless curves broken. In the first two scenarios, (a) and (b), there is at least one shearless curve in phase space preventing global transport. In the last scenario, (c), global transport takes place.

We performed a systematic study of these breakup scenarios by determining the shearless curve breakup parameter space that associates with each parameter value $(a,b,\epsilon)$ the corresponding shearless breakup scenario. Our results indicate a relation between periodic orbits reconnection-collision sequences and the breakup parameters of the central shearless curve, just as in the Standard Nontwist Map.

Besides the smooth boundaries, the parameter space also has regions with fractal boundaries, indicating a sensitive dependence of transport on the system parameters.

Moreover, even after all shearless curves are broken, a partial transport barrier persists in phase space. To investigate the strength of these partial barriers, we determined the transmissivity parameter space of the BNM. As transmissivity is the fraction of orbits that can overcome the barrier, the obtained results indicate that the strength of the partial barrier is sensitively dependent on the system parameters value.

However, there are some open questions about transport in the BNM. The behavior of the fractal dimension needs a more detailed study to analyze the difference between the dimension in the parameter spaces. Moreover, a deep study of the transport on BNM is needed, like analyzing the escape channels and manifold crossing.

\section*{Ackowledgments}
The authors thank the financial support from the Brazilian Federal Agencies (CNPq), grants \mbox{304616/2021-4} and 302665/2017-0, the S{\~a}o Paulo Research Foundation (FAPESP, Brazil) under grants 2018/03211-6, 2022/04251-7, 2022/08699-2 and 2022/05667-2, the Coordena\c{c}\~{a}o de Aperfei\c{c}oamento de Pessoal de N{\'i}vel Superior (CAPES) under Grant No. 88881.143103/2017-01, and the  Comit{\'e} Fran\c{c}ais d'Evaluation de la Coop{\'e}ration Universitaire et Scientifique avec le Br{\'e}sil (COFECUB) under Grant No. 40273QA-Ph908/18.

\appendix

\section{Symmetry properties of the BNM}
\label{sec:appendix1}

In this appendix, we review the symmetry properties of the Biquadratic Nontwist Map (BNM), as well as define and determine its indicator points. The BNM has spatial symmetry, i.e., let $M$ be the BNM and $S$ the transformation
\begin{equation}\label{eq:symmetry.transformation}
    S(x,y) = \left ( x + 1/2, \ -y \right ),
\end{equation}
\noindent the map $M$ is invariant under $S$, so, $M = S^{-1}MS$. Another property of the BNM is the time-reversal symmetry \cite{diego1996}, that is, we can decompose the map as a product of two involutions, in the way
\begin{equation}
    M = R_1R_0
\end{equation}
\noindent where
\begin{subequations}
\label{eq:involutions}
\begin{align}
R_0(x,y) &= \left ( -x, \ y - b \sin{(2\pi x)} \right),\\
R_1(x,y) &= \left (-x + a (1 - y^2)(1 - \epsilon y^2), \ y \right ).
\end{align}
\end{subequations}

Each involution \eqref{eq:involutions} has a set of invariant points, defined by
\begin{equation}
I_j = \left\{\mathbf{z} \ | \ R_j\mathbf{z} = \mathbf{z}\right\}, \ j=0,1,   
\end{equation}
\noindent which are one-dimensional sets called symmetry sets of the map. The set $I_0$ is formed by the union $s_1\cup s_2$, and  $I_1=s_3\cup s_4$, where $s_i$ is the $i$-th symmetry line given by
\begin{subequations}
\begin{align}
    s_1 &= \left\{ \ (x,y) \ | \ x = 0 \ \right\},\\
    s_2 &= \left\{ \ (x,y) \ | \ x = 1/2 \ \right\},\\
    s_3 &= \left\{ \  (x,y) \ | \ x = a (1-y^2)(1-\epsilon y^2)/2 \ \right\},\\
    s_4 &= \left\{ \ (x,y) \ | \ x = a (1-y^2)(1-\epsilon y^2)/2 + 0.5 \ \right\}.
\end{align}
\end{subequations}

As the standard nontwist map, the BNM central shearless curve is invariant under the symmetry transformation $S$, and by the involutions $R_j$. Shearless invariant curves that have those properties have indicator points, defined as fixed points of the transformation $SR_j$, that is,
\begin{equation}
    SR_j(x,y) = (x,y), \ \ j=0,1,
\end{equation}
which belong to the shearless curve, if it exists \cite{shinohara1997}. Performing the calculation for the BNM, we obtain the indicator points
\begin{equation}
    \mathbf{P}_0^\pm =\left(\pm \dfrac{1}{4}, \pm \dfrac{b}{2} \right) \ \ \text{and} \ \ \mathbf{P}_1^\pm =\left( \dfrac{a}{2} \pm \dfrac{1}{4}, 0 \right),
\end{equation}
which are exactly the same as the SNM indicator points.

\bibliographystyle{plain}
\bibliography{output.bbl}
	
\end{document}